\documentclass[twocolumn,floatfix,prl]{revtex4}%
\usepackage[dvipdfmx]{graphicx}%
\usepackage{amsmath}%
\usepackage{amsfonts}
\usepackage{amssymb}
\usepackage{hhline}
\usepackage{bm}
\usepackage{mathrsfs}
\usepackage{color}

\def\s{{\sigma}}

\def\k{{ {\bm k} }}
\def\p{{ {\bm p} }}
\def\q{{ {\bm q} }}
\def\Q{{ {\bm Q} }}

\def\0{{ {\bm 0} }}
\def\w{{\omega}}
\def\a{{\alpha}}
\def\b{{\beta}}

\allowdisplaybreaks[4]

\begin{document}
\title{
Multipole fluctuation theory for heavy fermion systems: \\
Application to multipole orders in CeB$_6$
}
\author{
Rina Tazai and Hiroshi Kontani
}

\date{\today }

\begin{abstract}
In heavy fermion systems, 
multipole degrees of freedom make possible 
the emergence of rich phenomena,
such as hidden orders and superconductivities.
However, many of them remain unsolved since the 
origin of higher-rank multipole interaction is not well understood.
Among these issues,
we focus on the quadrupole order in CeB$_6$,
which is a famous multipolar heavy fermion system
actively studied for decades.
We analyze the multiorbital periodic Anderson model for CeB$_6$,
and find that both magnetic, quadrupole, and octupole fluctuations
develop cooperatively due to the strong inter-multipole coupling
given by higher-order many-body effects, called the vertex corrections.
It is found that the
antiferro-quadrupole order in CeB$_6$ is driven by 
the interference between magnetic-multipole fluctuations.
The discovered inter-multipole coupling
mechanism is a potential origin of various hidden orders 
in various heavy fermion systems.

\end{abstract}

\address{
Department of Physics, Nagoya University,
Furo-cho, Nagoya 464-8602, Japan. 
}
 

\sloppy

\maketitle

Heavy fermion (HF) systems are very interesting 
platform of exotic electronic states induced by
strong Coulomb interaction and spin-orbit interaction (SOI)
on $f$-electrons.
Magnetic fluctuations cause interesting 
quantum critical phenomena and superconductivity
\cite{Coleman,Moriya,Yamada,Kontani-rev,Monthoux,Tremblay,Scalapino,Takimoto-SC}.
In addition, higher-rank multipole operators are also active thanks to the
strong SOI of $f$-electrons.
For this reason, various interesting multipole order and fluctuations,
which are absent in transition metal oxides, emerge in HF systems.
One of the most famous example is the multipole-order in CeB$_6$:
The antiferro-quadrupole order with $\q=(\pi,\pi,\pi)$
occurs at $T_Q=3.2$K, and magnetic order appears at $T_N=2.4$K
\cite{Kasuya,Goto,Sera,Inosov-review}.
In addition, antiferro-octupole order is stabilized under 
weak magnetic field
\cite{Shiina1,Shiina2,Shiina3,Shiina4}.
Thus, various ranks of multipole orders appear simultaneously
in the phase diagram of CeB$_6$.
This fact indicates that different multipoles are strongly entangled,
which would be universal in HF system.


Up to now, multipole orders in CeB$_6$ has been studied 
actively based on the localized $f$-multipole models
\cite{Shiina1,Shiina2,Shiina3,Shiina4,Sera2,Kusunose,Hanzawa}.
However, recent ARPES studies 
\cite{ARPES1,ARPES2}
revealed that the $f$-electron is itinerant for $T\sim T_Q$. 
The characteristic dynamical magnetic 
susceptibility in CeB$_6$ measured by neutron inelastic scattering 
\cite{Inosov1,Inosov2} 
is explained in the itinerant picture
based on the periodic Anderson model (PAM)
\cite{Thal}.
If we apply the random-phase-approximation (RPA) for the PAM, 
however, quadrupole order cannot be obtained.
In fact, only odd-rank (=magnetic) multipole fluctuations develop,
whereas even-rank (=electric) multipole ones remain small in the RPA
\cite{Thal,Ikeda-U,Tazai-HF}.
This fact means that the importance of vertex corrections (VCs),
which represent the many-body effects ignored in the RPA.
The lowest-order VC with respect to fluctuations,
called the Maki-Thompson (MT) type VC,
gives the rank-5 multipole order in URu$_2$Si$_2$ \cite{Ikeda-U}.
However, the MT-VC does not magnify the even-rank multipole fluctuations.
Thus, microscopic origin of quadrupole order,
which frequently appears in various compounds, 
is still unsolved.
CeB$_6$ is a suitable platform to 
construct a theory of multipole order in HF systems.


In Fe-based and cuprate superconductors,
significant roles of the Aslamazov-Larkin (AL) VC,
which is the second-order VC with respect to fluctuations,
attract considerable attention
\cite{Onari-SCVC,Onari-FeSe,Yamakawa-FeSe}.
The AL-VC describes various spin-fluctuation-driven
nematicities, such as orbital order and bond order,
that fail to be explained by the RPA.
The significance of the AL-VC near the magnetic 
criticality is confirmed 
by different theoretical studies, especially 
by the functional-renormalization-group (fRG) studies
\cite{Tsuchiizu1,Tsuchiizu2,Tsuchiizu3,Tazai-RG,Tsuchiizu4,Chubukov-RG1,Schmalian}.
In HF systems,
phonon-mediated superconductivity
can be stabilized by the AL-process for the electron-boson coupling.
This mechanism may be responsible for the fully gapped 
$s$-wave state in CeCu$_2$Si$_2$
\cite{Tazai-HF,Tazai-JPSJ}.
These findings
indicate that the AL-VC plays essential roles in HF systems

In this paper, 
we study the mechanism of quadrupole order in CeB$_6$
based on the itinerant $f$-electron picture,
by considering the AL-VC for multipole susceptibilities.
For this purpose, we introduced an effective PAM for CeB$_6$
with $\Gamma_8$ quartet $f$-orbital basis.
Both ferro- and antiferro-magnetic and octupole fluctuations 
are induced by the Fermi surface nesting,
consistently with recent neutron experiments.
Then, antiferro-quadrupole ($O_{xy}$) order 
is induced by the interference between different 
magnetic multipole fluctuations.
The present multipole fluctuation theory with introducing AL-VC
will be applicable for various HF systems.

In HF systems, the DMFT has been applied successfully
\cite{LDADMFT_multiporbital,KotVol-DMFT,Held,DMFT-HF,Otsuki-HF}.
In the DMFT, the irreducible VC is local.
Here, we calculate the $\k$-dependence (=nonlocality) 
of AL and MT diagrams accurately in order to evaluate the 
interference between multipole fluctuations.


Here, we introduce a two-dimensional 
PAM as an effective model for CeB$_6$.
For $f$-electron states,
we consider the $\Gamma_{8}$ quartet in $J=5/2$ space
\cite{Shiina1}:
%
\begin{eqnarray}
|f_{1}\Sigma \rangle=\sqrt{\frac{5}{6}}|\mp \frac{5}{2}\rangle+\sqrt{\frac{1}{6}}|\pm\frac{3}{2}\rangle, && |f_{2}\Sigma \rangle=|\mp \frac{1}{2}\rangle, 
\label{eqn:wavefunc}
\end{eqnarray}
where $\Sigma=\pm$ is the pseudo-spin of $f_{l}$-orbital ($l=1,2$).
The kinetic term is given by
$\hat{H}_{0}=\sum_{\k\sigma}\epsilon_{\k}c^{\dagger}_{\k\sigma}c_{\k\sigma}+\sum_{\k l\sigma}E_{f}f^{\dagger}_{\k l\sigma}f_{\k l\sigma}
+\sum_{\k l\sigma}
\left(V^{*}_{\k l\sigma}f^{\dagger}_{\k l\sigma}c_{\k\sigma}+{\rm h.c}\right)$,
%
where $c^{\dagger}_{\k \sigma}$ is a creation operator 
for $s$-electron with momentum $\k$ and spin $\sigma$ on Ce ion.
$\epsilon_{\k}$ is the conduction band dispersion,
which we explain in the Supplemental Material (SM) A
\cite{SM}.
$f^{\dagger}_{\k l \Sigma}$ is a creation operator for 
$f$-electron with $\k$, orbital $l$ and pseudo-spin $\Sigma$.
$V_{\k l\sigma}$ is the $s$-$f$ hybridization term 
between the nearest Ce cites.
In the two-dimensional model,
the pseudo-spin and $s$-electron spin are conserved 
($\sigma=\Sigma$) in the $s$-$f$ mixing \cite{Tazai-HF}.
Using the tight-binding method 
\cite{Takegahara}, 
$V_{\k l\sigma}$ is given as
\begin{eqnarray}
V_{\k f_{l}\uparrow}&=&-A_{l}t_{sf}(\sin k_{y} +(-1)^l i\sin k_{x} ),
\label{eqn:hybri}
\end{eqnarray}
and $V_{\k f_{l}\downarrow}= -V_{\k f_{l}\uparrow}^{*}$.
We set $A_l=\sqrt{{18}/{14}}$,
and give a detailed explanation on $V_{\k l\sigma}$
in the SM A \cite{SM}.
Hereafter, we set $2|t_{ss}^{1}|=1$ as energy unit,
and put $t_{sf}=0.78$, $E_{f}=-2.0$, $T=0.01$, and $\mu=-2.45$.
Then, $f(s)$-electron number is $n_{f}=0.58$ ($n_{s}=0.69$).

\begin{figure}[htb]
\includegraphics[width=.85\linewidth]{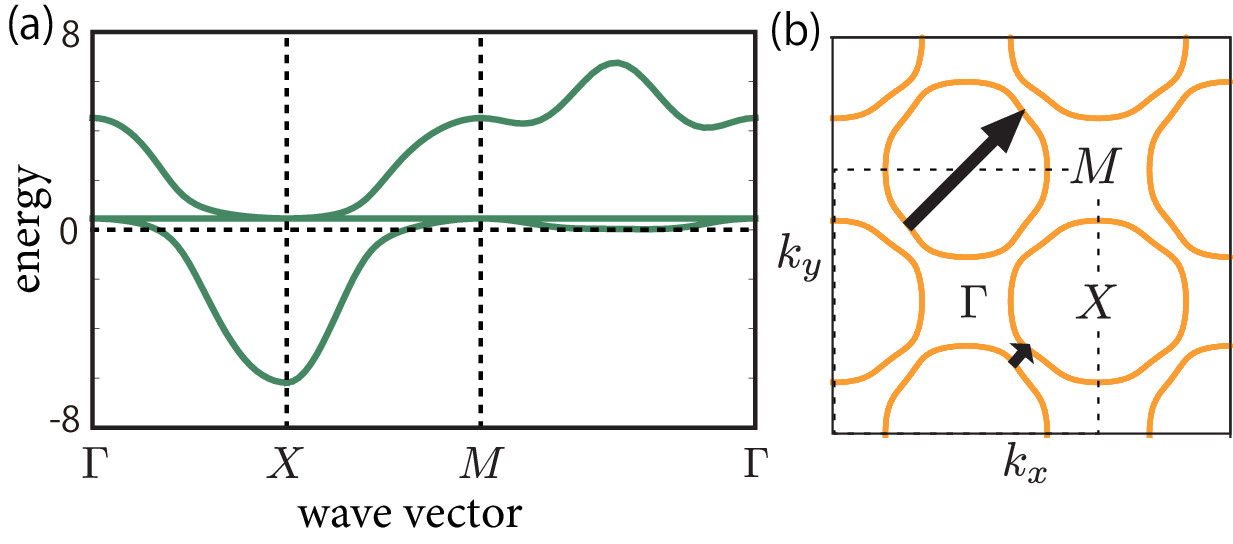}
\caption{
(a) Band dispersion and (b) Fermi surfaces
of the present model.
Major nesting vectors are shown.
}
\label{fig:band}
\end{figure}

Figure \ref{fig:band}(a) shows the band structure of PAM.
The lowest band crosses the Fermi level ($\epsilon=0$).
Since $W_{D}\sim 5$eV in CeB$_{6}$ \cite{ARPES1,ARPES2},
$2|t_{ss}^{1}|(=1)$ corresponds to $\sim0.5$eV.  
The bandwidth of itinerant $f$-electron is 
$W_{D}^{qp} \sim |V| \sim 1$.
The Fermi surfaces shown in Fig. \ref{fig:band}(b) 
are composed of large ellipsoid electron pockets around X,Y points, 
consistently with recent ARPES studies
\cite{ARPES1,ARPES2}.

We also introduce the Coulomb interaction term
$\hat{H}_{U}=u \hat{H}^{0}_{U}$.
Here, $\hat{H}^{0}_{U}=\frac14 \sum_{LL'MM'} U^0_{L,L';M,M'}
f_{L}^\dagger f_{L'} f_{M} f_{M'}^\dagger$,
where $L=(l,\sigma)$ and $M=(m,\rho)$.
$\hat{U}^0$ is the normalized Coulomb interaction for Ce-ion;
the maximum element of $\hat{U}^0$ is set to unity. 
The detailed explanation is given in Ref. \cite{Tazai-HF}
and in the SM A \cite{SM}.


In the present $\Gamma_{8}$ quartet model, there are 
16-type active multipole operators up to rank 3; monopole, 
dipole (rank 1), quadrupole (rank 2), octupole (rank 3) momenta.
The table of irreducible representation (IR)
for the $D_{4h}$ two-dimensional model 
is shown in TABLE {\ref{tab:multipole} \cite{Ikeda-U}.
An even-rank (odd-rank) operator corresponds to 
an electric (magnetic) multipole operator.
The $4\times4$ matrix form of each operator, ${\hat Q}$,
is shown in the SM B \cite{SM}.
%
\begin{table}[htb]
  \begin{tabular}{|c|c|c|c|} \hline
    \hspace{3mm}IR ($\Gamma$) \hspace{3mm} & rank (k) & Operator ($\hat{Q}$) & IR in $H_{z}$
     \\ \hhline{|=|=|=|=|}
     $\Gamma_{1}^{+}$ & $0$  & $\hat{1}$       & $\Gamma_{1}$ \\ \cline{2-3}
                             & $2$  & $\hat{O}_{20}$ &  \\ \hline
     $\Gamma_{3}^{+}$ & $2$  & $\hat{O}_{22}$ & $\Gamma_{3}$ \\ \hline
     $\Gamma_{4}^{+}$ & $2$  & $\hat{O}_{xy}$ & $\Gamma_{4}$\\   \hline
     $\Gamma_{5}^{+}$ & $2$  & $\hat{O}_{yz},\hat{O}_{zx}$ & $\Gamma_{5}$\\  \hhline{|=|=|=|=|}
     $\Gamma_{2}^{-}$ & $1$  & $\hat{J}_{z}$          & $\Gamma_{1}$  \\  \cline{2-3}
                             & $3$  & $\hat{T}_{z\a}$ &  \\   \hline
     $\Gamma_{3}^{-}$ & $3$  & $\hat{T}_{xyz}$ & $\Gamma_{4}$ \\  \hline
     $\Gamma_{4}^{-}$ & $3$  & $\hat{T}_{z\b}$ & $\Gamma_{3}$ \\ \hline
     $\Gamma_{5}^{-}$ & $1$  & $\hat{J}_{x}$,$\hat{J}_{y}$ & $\Gamma_{5}$ \\  \cline{2-3}
                             & $3$  & $\hat{T}_{x\a}$,$\hat{T}_{y\a}$ &   \\  \cline{2-3}
                             & $3$  & $\hat{T}_{x\b}$,$\hat{T}_{y\b}$ &  \\ \hline
    \end{tabular}
\caption{IRs and 16-type active multipole operators of $D_{4h}$ point group. 
Operator with rank $k$ corresponds to 
$2^{k}$-pole.}
    \label{tab:multipole}
\end{table}

Here, we calculate the $f$-electron susceptibility.
The bare irreducible susceptibility is given by
$\chi_{\a,\b}^{0}(q)= -T\sum_{k}G^{f}_{LM}(k+q)G^{f}_{M' L'}(k)$, 
where  $q\equiv (\q, \omega_{n})=(\q,2j\pi T)$, 
$\a\equiv (L,L')$ and $\b\equiv (M,M')$.
Here, $\a,\b$ takes $1\sim 16$, and $\hat{G}^f$
is the Green function without self-energy \cite{Tazai-HF}.
We also consider the VCs due to AL and MT terms, 
$\hat{X}^{\rm{AL+MT}}$, which we will explain later.
Then, $f$-electron susceptibility is given as
\begin{eqnarray}
\hat{\chi}(q)= {\hat \phi}(q)
({\hat 1}-u{\hat U}^0{\hat \phi}(q))^{-1},
\label{eq:rpa}
\end{eqnarray}
where $\hat{\phi}(q)=\hat{\chi}^{0}(q)+\hat{X}^{\rm{AL+MT}}(q)$ 
is irreducible susceptibility including the VCs 
in the $16\times 16$ matrix form.

Here, we consider the following eigen equation
\begin{eqnarray}
u\hat{U}^0\hat{\phi}(\q,0)\vec{w}^{\Gamma}(\q)
= \a^\Gamma(\q)\vec{w}^{\Gamma}(\q).
\label{eqn:eigenequation}
\end{eqnarray}
When the eigenvector is expressed as
$\vec{w}^{\Gamma}(\q)=\sum_{Q\in\Gamma}Z^{Q}(\q)\vec{Q}$,
the maximum of the eigenvalue $\a^\Gamma(\q)$ gives 
the Stoner factor for IR $\Gamma$, 
$\a^\Gamma =\max_\q\{\a^\Gamma(\q)\}$.
Here, $\vec{Q}$ is $16\times 1$ vector
defined as $(\vec{Q})_\a= (\hat{Q})_{L,L'}$
and $Z^{Q}(\q)$ is a real coefficient.
The $\Gamma$-channel multipole order appears
when $\a^\Gamma\ge1$.
The inner product 
$(\vec{Q})^{\dagger}\vec{Q}'$ is unity for $Q=Q'$.
It is zero when $Q$ and $Q'$ belong to different IR,
whereas it is not always zero when $Q\ne Q'$ belong to 
the same IR \cite{Tazai-HF,SM}.
We introduce the magnetic (electric) Stoner factor as 
$\a^{\rm mag(el)}=\max_{n}\{\a^{\Gamma_n^{-(+)}}\}$.

\begin{figure}[htb]
\includegraphics[width=.79\linewidth]{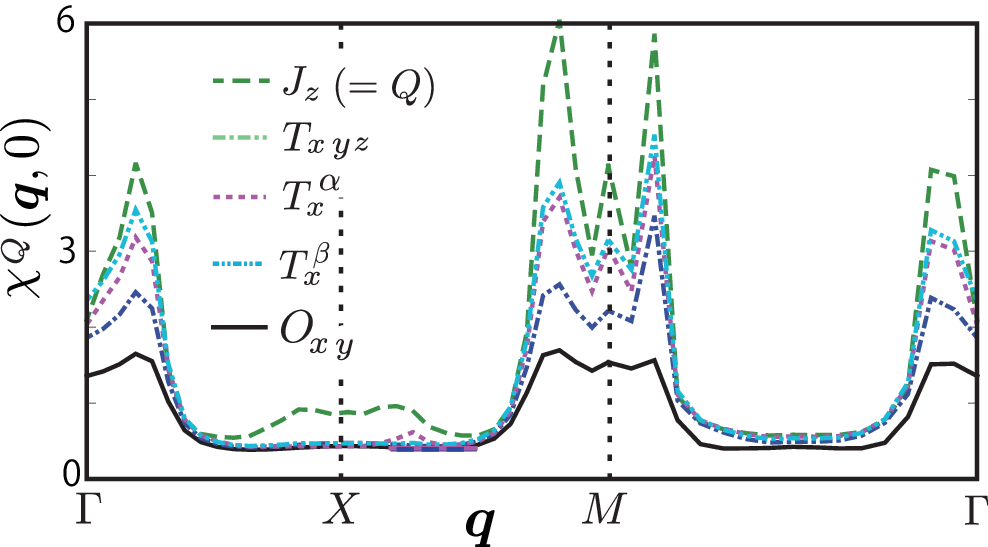}
\caption{
Obtained multipole susceptibilities by the RPA.
The peak positions correspond to the nesting vectors 
in Fig. \ref{fig:band} (b).
}
\label{fig:chispin}
\end{figure}

Using $\vec{Q}$,
the multipole susceptibility is given by
\begin{eqnarray}
\chi^{Q,Q'}(q)=(\vec{Q})^{\dagger}\hat{\chi}(q)\vec{Q}'.
\label{eq:suscep}
\end{eqnarray} 
%
First, we show the numerical results by the RPA,
given as $X^{\rm{AL+MT}}=0$.
Figure \ref{fig:chispin} shows obtained susceptibilities
$\chi^{Q}(\q,0)\equiv \chi^{Q,Q}(\q,0)$ at $u=1.08$
($\alpha^{\rm mag}=0.9$).
In the RPA, $\chi^{J_{z}}$ is the most largest.
Secondly, $\chi^{T_{\nu}^{\b}}, \chi^{T_{\nu}^{\a}}(\nu=x,y)$
and $\chi^{T_{xyz}} $ are also enlarged.
$\chi^{J_{z}}(\q,0)$ has peak value at $\q\approx \bm{0}$ 
and $\q\approx\bm{Q}\equiv (\pi,\pi)$,
which is consistent with the inelastic neutron-scattering
that reports strong ferromagnetic and antiferromagnetic 
($\q=(\pi,\pi,\pi),(\pi,\pi,0)$) fluctuations above $T_N$
\cite{Inosov2,comment2}.
Therefore, the present two-dimensional PAM is reliable.

On the other hand, the RPA quadrupole susceptibility
remains small.
To understand this result, 
we examine the $(Q,Q')$ component of normalized Coulomb interaction:
%
\begin{eqnarray}
U^{Q,Q'}_0=(\vec{Q})^\dagger \hat{U}^0 \vec{Q}' .
\end{eqnarray}
TABLE \ref{tab:multipole2} shows the diagonal component
$U^{Q}_0\equiv U^{Q,Q}_0$.
Since $U^{Q}_0$ for the quadrupole channels
is much smaller than that for the dipole and octupole channels,
the quadrupole susceptibilities is small within the RPA.

\begin{table}[htb]
  \begin{tabular}{|c|c|c|c|c|c|c|c|} \hline
  Q &   1 & $O_{20(22)}$  & $O_{xy(yz,zx)}$ & $T_{xyz}$ & $J_{z(x,y)}$ &  $T^{\alpha}_{z(x,y)}$ &$T^{\beta}_{z(x,y)}$
     \\ \hhline{|=|=|=|=|=|=|=|=|}
 $U^{Q}_0$ &   -2.4 & 0.50  & 0.63 & 0.81 & 1.03 & 0.94 & 0.94 \\ \hline
   \end{tabular}
\caption{Normalized Coulomb interaction $U^{Q}_0$.
$U^{Q,Q'}_0=0$ for $Q\ne Q'$ except for $U^{J_{\mu},T^{\a}_{\mu}}_0=0.58$ ($\mu=x,y,z$). }
\label{tab:multipole2}
\end{table}

From now on, we introduce the VCs due to AL and MT terms.
Diagrams of these VCs are shown in Fig.\ref{fig:chiorb} (a).
For example, the AL1 term is given as
\begin{eqnarray}
X^{\rm{AL1}}_{\a \b}(q)=\frac{T}{2}\sum_{\a' \a'' \b' \b''}
\Lambda_{\a' \b''}^{\a} (q,p) V_{\a' \b'} (p-q) \nonumber \\
\times  V_{\a'' \b''}(p) \Lambda_{\b' \a''}^{\b *} (\bar{q},\bar{p}),
\label{eqn:UALc}
\end{eqnarray}
%
%
where $p\equiv (\p, \omega_{m})$, $\bar{p}\equiv (\p,-\omega_{m})$, 
and
$\hat{V}(q)\equiv u^{2}\hat{U}^0\hat{\chi}(q)\hat{U}^0+u\hat{U}^0$
is the dressed interaction given by the RPA.
The three-point vertex is given as
\begin{eqnarray}
\Lambda^{EF}_{ABCD} (q,p) \equiv -T\sum_{k}G^{f}_{AF}(k-q)G^{f}_{EC}(k)
G^{f}_{DB}(k-p).
\end{eqnarray}
Other VCs are explained in the SM C \cite{SM}.

Figures \ref{fig:chiorb} (b) and (c) show the 
obtained quadrupole susceptibility by including 
MT- and AL-VCs.
In contrast to the RPA result,
the obtained $\chi^{O_{xy}}(\q,0)$ is strongly enhanced 
at $\q=\bm{Q}$ and $\q=\bm{0}$,
and becomes the largest of all $\chi^{Q}$.
This enhancement originates from the AL terms, 
whereas the MT term is very small
as we show in SM C \cite{SM}.
The obtained $\chi^{O_{xy}}(\q,0)$ 
has the highest peak at $\q=\bm{Q}$, consistently with 
the antiferro-$O_{xy}$ order in CeB$_6$.
Moreover, the second highest peak of $\chi^{O_{xy}}(\q,0)$ 
at $\q=\bm{0}$ explains the softening of shear modulus $C_{44}$ 
in CeB$_6$ \cite{Goto}.
We show other quadrupole susceptibilities 
in the SM C \cite{SM}.
To summarize, the obtained strong enhancements of 
$\chi^{O_{xy}}(\q,0)$ and $\chi^{J_z}(\q,0)$ 
at both $\q=\bm{Q}$ and $\q=\bm{0}$
reproduce the key experimental results of CeB$_6$.

\begin{figure}[htb]
\includegraphics[width=.85\linewidth]{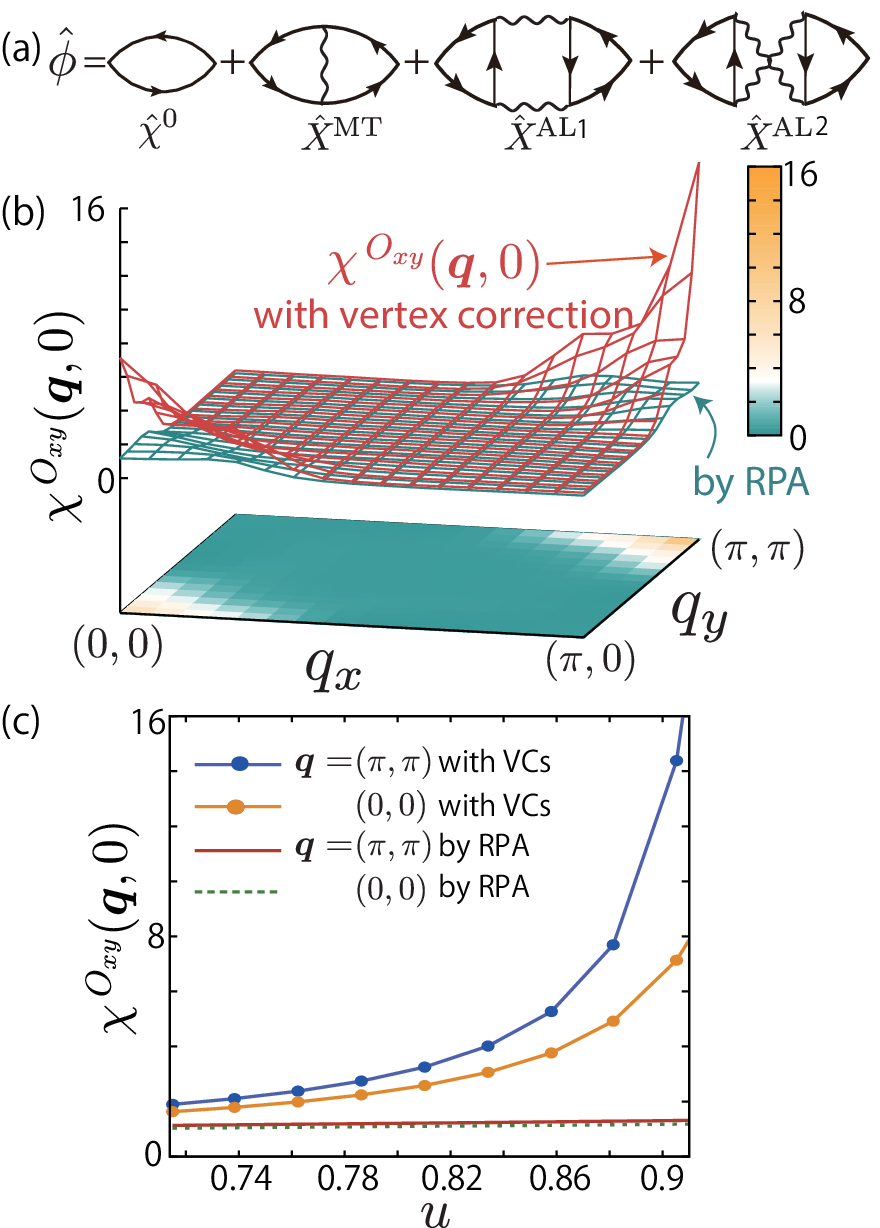}
\caption{
(a) Diagrams of the irreducible susceptibility 
$\hat{\phi}$ with MT- and AL-VCs.
(b) $\q$-dependence of $\chi^{O_{xy}}(\q,0)$;
$\a^{\Gamma_4^{+}}=0.94$ with VCs. 
(c) $u$-dependence of  $\chi^{O_{xy}}(\q,0)$ at $\q=\bm{Q},\bm{0}$.}
\label{fig:chiorb}
\end{figure}

Next, we explain that the $O_{xy}$ quadrupole order
is derived from the interference between magnetic multipole fluctuations.
For this purpose, 
we analyze the total AL term 
$\hat{X}\equiv \hat{X}^{\rm AL1}+\hat{X}^{\rm AL2}$
for $O_{xy}$-channel defined as
\begin{eqnarray}
X_{O_{xy}}(q)\equiv (\vec{O}_{xy})^{\dagger} \hat{X}(q) \vec{O}_{xy} ,
\label{eqn:VQQ}
\end{eqnarray}
where $\hat{X}\equiv \hat{X}^{\rm AL1}+\hat{X}^{\rm AL2}$.
The Stoner factor for $O_{xy}(=\Gamma_4^+)$ channel is
proportional to $u U^{O_{xy}}_0 \phi_{O_{xy}}(q)$, where 
$\phi_{O_{xy}}(q)\equiv (\vec{O}_{xy})^{\dagger} \hat{\phi}(q)\vec{O}_{xy}$.
Therefore, $X_{O_{xy}}(q)\ (>0)$ 
works as enhancement factor of $O_{xy}$ susceptibility.

By following Ref. \cite{Tazai-HF},
we expand $\hat{V}(q)$ on the basis of multipole 
operator as 
\begin{eqnarray}
\hat{V}(q)=\sum_{QQ'}v^{QQ'}_{q}\vec{Q}(\vec{Q}')^{\dagger},
\label{eqn:VQQ2}
\end{eqnarray}
where the real coefficient $v^{QQ'}_q$ is uniquely determined
\cite{Tazai-HF}.
From Eq.(\ref{eqn:UALc}), (\ref{eqn:VQQ}) and (\ref{eqn:VQQ2}), 
the AL1 term due to $(Q,Q')$-channel fluctuations is given as
\begin{eqnarray} 
X_{O_{xy}}^{{\rm AL1},Q Q'}(q)\equiv \frac{T}{2}
\sum_{p}v^{Q}_{p} v^{Q'}_{p-q} \Lambda_{q,p}^{O_{xy}QQ'} 
(\Lambda_{\bar{q},\bar{p}}^{O_{xy}Q'Q})^{*} ,
\label{eqn:XQQ}
\end{eqnarray} 
where $v^{Q} \equiv v^{QQ}$ and 
$\Lambda_{q,p}^{O_{xy}QQ'}$ is defined as
\begin{eqnarray}
\Lambda_{q,p}^{O_{xy}QQ'} \equiv \sum_{\a} (\vec{O}_{xy})^{*}_\a 
(\vec{Q}')^{\dagger} \hat{\Lambda}^{\a}(q,p) \vec{Q}.
\label{eqn:LQQ}
\end{eqnarray} 
The diagrammatic expression of Eq. (\ref{eqn:XQQ})
is shown in Fig.\ref{fig:kaiseki}(a).
Figure \ref{fig:kaiseki}(b) shows the $\q$-dependence of 
$X_{O_{xy}}^{Q Q'}(\q,0)$ at $u=0.91$.
We find that the $(Q,Q')=(T_{x}^{\a},T_{y}^{\a})$,
$(J_{z},T_{xyz})$, $(T_{x}^{\b},T_{y}^{\b})$
channels give the dominant contributions.
Other terms not shown in Fig.\ref{fig:kaiseki}(b)
give negligible contribution.

\begin{figure}[htb]
\includegraphics[width=.85\linewidth]{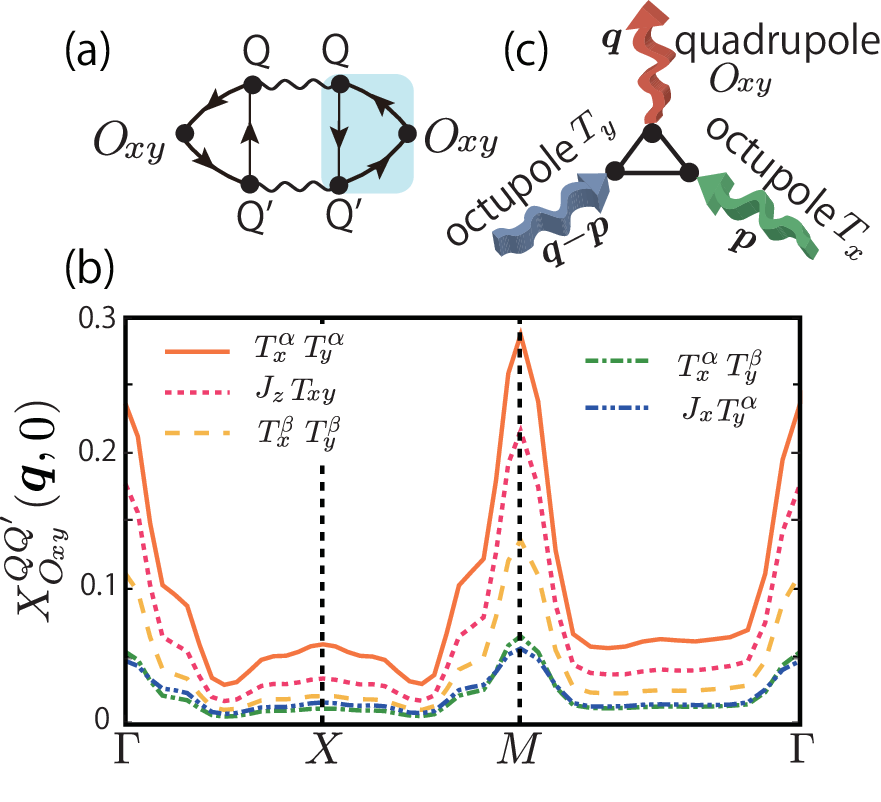}
\caption{
(a) AL-term $X_{O_{xy}}^{{\rm AL1},Q Q'}$
given by $(Q,Q')$-channel fluctuations.
(b) Obtained $X_{O_{xy}}^{Q Q'}(\q,0)$
(c) Quantum process of $O_{xy}$ fluctuations driven by 
the interference between $(T_{x},T_{y})$ fluctuations,
which corresponds to the shaded area in (a).
}
\label{fig:kaiseki}
\end{figure}

Figure \ref{fig:kaiseki}(c) presents the quantum process of 
$O_{xy}$ quadrupole order driven by 
the interference between $(T_{x},T_{y})$ fluctuations,
which corresponds to $\Lambda^{O_{zx}T_xT_y}$ 
in Fig.\ref{fig:kaiseki}(a).
This process is realized when
$\Lambda^{O_{zx}QQ'}\sim
{\rm Tr}\{ \hat{O}_{xy}\cdot \hat{Q}\cdot \hat{Q}' \}\ne 0$.
Since $\Lambda^{QTT'}=0$ for odd-rank $Q$, 
the AL-VC is unimportant for $\chi^J$ and $\chi^T$
 \cite{Yamakawa-FeSe}.

Next, the $\q$-dependence of the AL-VC is given as
$X_{O_{xy}}^{T_x T_y}(\q)\propto \sum_{\p}\chi^{T_x}(\p) \chi^{T_y}(\q-\p)$,
which becomes large at $\q=\bm{Q}$ and $\q=\bm{0}$
since $\chi^{T_\mu}(\p)$ has large peaks at $\p\sim\bm{Q},\bm{0}$
shown in Fig. \ref{fig:chispin}.
Thus, antiferro-quadrupole order in CeB$_6$
originates from the interference between ferro- and 
antiferro-magnetic multipole fluctuations.

Finally, we discuss the field-induced octupole order,
which has been studied intensively as a main issue of CeB$_6$
\cite{Shiina1,Shiina2,Shiina3,Shiina4}.
The Zeeman term under the magnetic field along $z$-axis 
is given as
$\hat{H}_Z=h_{z}\sum_{L,M}(\hat{J}_{z})_{L,M} f^{\dagger}_{\k L} f_{\k M}$.
When $h_{z}\ne0$,
both $O_{xy}$ and $T_{xyz}$ belong to the same IR $\Gamma_{4}$
shown in TABLE \ref{tab:multipole}
\cite{Shiina1}. 
Therefore, large quadrupole-octupole susceptibility
$\chi^{O_{xy},T_{xyz}}(\q,0)$ is induced in proportion to $h_z$.
To verify this, 
we solve the eigen equation (\ref{eqn:eigenequation})
for the IR $\Gamma_4$ under $h_z$,
at the fixed magnetic Stoner factor in the RPA $\a^{\rm mag}=0.8$
\cite{comment,Sakurazawa}.

Figures \ref{fig:ziba}(a) and (b) show the obtained eigenvector
$\vec{w}^{\Gamma_4}(\q)=Z^{O_{xy}}(\q)\vec{O}_{xy} + Z^{T_{xyz}}(\q)\vec{T}_{xyz}$ 
($|\vec{w}^{\Gamma_4}|^2=1$) 
and the Stoner factor $\a^{\Gamma_4}$ at $\q=\bm{Q}$,
respectively, as functions of $h_z$.
Here, $\a^{\Gamma_4}$ is the largest Stoner factor.
The increment of $\a^{\Gamma_4}$ under $h_z$ is 
consistent with the field-enhancement of $T_Q$ in CeB$_6$.
(In contrast, $T_N$ will be suppressed by large $O_{xz}$ moment.)
Also, $Z^{T_{xyz}}$ increases linearly in $h_z$,
due to the interference process under $h_z$
shown in the inset of Fig. \ref{fig:ziba}(b).
$Z^{T_{xyz}}$ becomes comparable to $Z^{O_{xy}}$ 
under small magnetic field $h_{z} \lesssim 0.03 \ll W_D^{qp}/10$. 
Since the ratio of the ordered momenta at $T_Q$ is 
$M^{T_{xyz}}/M^{O_{xy}}=Z^{T_{xyz}}/Z^{O_{xy}}$,
field-induced antiferro-$T_{xyz}$ order 
is naturally explained.
 

\begin{figure}[htb]
\includegraphics[width=.8\linewidth]{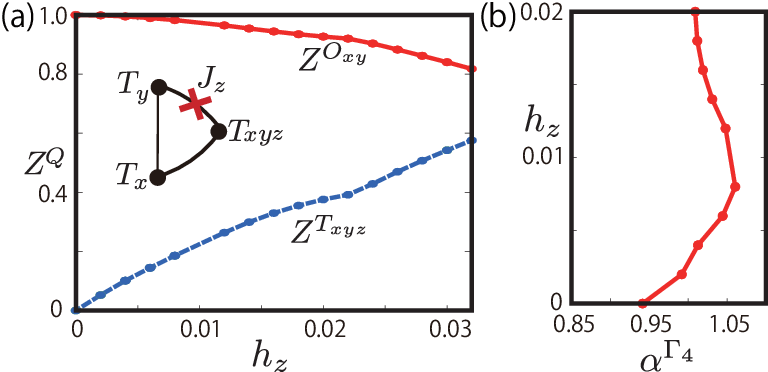}
\caption{
(a) Form factor $(Z^{O_{xy}},Z^{T_{xyz}})$ of the eigenvector
for $\Gamma_4= \{O_{xy},T_{xyz}\}$ at $\q=\bm{Q}$ under $h_z$.
Inset: $h_z$-linear term of the three-point vertex
$\Lambda^{T_{xyz}T_xT_y}$ that gives large $\chi^{O_{xy}T_{xyz}}(\q,0)$.
(b) Stoner factor $\a^{\Gamma_4}$ as function of $h_z$.
}
\label{fig:ziba}
\end{figure}

In summary,
we developed multipole fluctuation theory by 
focusing on the AL-type VCs in HF systems,
and applied the theory to the multipole order physics in CeB$_6$.
Both ferro- and antiferro-magnetic multipole fluctuations 
emerge in CeB$_6$ due to the nesting of Fermi surfaces,
consistently with neutron experiments.
Then, antiferro-$O_{xy}$ order in CeB$_6$ at $T_Q\ (>T_N)$
is derived from the interference between 
different magnetic multipole fluctuations,
which is depicted in Fig. \ref{fig:kaiseki} (c).
We also explained the field-induced octupole order,
which is a central issue of CeB$_6$.
The discovered inter-multipole coupling mechanism
will be significant in various HF systems,
such as quadrupole ordering system
Pr$T_2$Zn$_{20}$ ($T$ = Rh and Ir) \cite{Oni-Pr}
and Pr$T_2$Al$_{20}$ ($T$=V,Ti) \cite{Naka-Pr}.
Although the analysis of AL-VC in three-dimensional PAM is very difficult,
it is an important future problem.

We stress that 
the on-site quadrupole ($O_{xy}$) interaction on Ce-ion
is about 60\% of dipole ($J_\mu$) one 
as shown in TABLE \ref{tab:multipole2}.
Therefore, quadrupole order cannot appear 
within the mean-field theory.
In contrast, in the localized RKKY model,
quadrupole interaction is as large as
the dipole interaction
\cite{Shiina1,Shiina4}.
Such discrepancy between 
itinerant picture and localized one,
which is an important problem in HF systems,
is partially resolved by considering the VCs as we discussed here.


\acknowledgements
We are grateful to S. Onari and Y. Yamakawa for useful discussions.
This study has been supported by Grants-in-Aid for Scientific
Research from MEXT of Japan.




\clearpage

\makeatletter
\renewcommand{\thefigure}{S\arabic{figure}}
\renewcommand{\theequation}{S\arabic{equation}}
\makeatother
\setcounter{figure}{0}
\setcounter{equation}{0}
\setcounter{page}{1}
\setcounter{section}{1}

\begin{widetext}
\begin{center}
{\bf 
[Supplementary Material] \\
Multipole fluctuation theory for heavy fermion systems: 
Application to multipole orders in CeB$_6$
}%
\end{center}

\begin{center}
Rina Tazai and Hiroshi Kontani
\end{center}

\begin{center}
\textit{Department of Physics, Nagoya University, Nagoya 464-8602, Japan}
\end{center}

\end{widetext}
\subsection{A: model Hamiltonian}

Here, we present detailed explanation for the 
model Hamiltonian.
In CeB$_6$, the conduction band is composed of
$5d$ electrons on Ce-ions,
Here, to simplify the model Hamiltonian,
we introduce the conduction band made of $s$ electrons.
The realistic tight-binding model of conduction band
of CeB$_6$ is given in Ref. \cite{S-ARPES2}.
In the present study, we slightly modify the model 
in Ref. \cite{S-ARPES2} and put $k_z=0$,
in order to reproduce the 
experimental Fermi surfaces of CeB$_6$ on the 
$k_x$-$k_y$ plane after $s$-$f$ hybridization.
The present two-dimensional tight-binding model 
for conduction band is given as 
\begin{eqnarray}
\epsilon_{\k}&=&t_{ss}^{1}\left( \cos k_{x}+\cos k_{y} \right) \nonumber \\ 
&+&t_{ss}^{2}\left\{ \cos(k_{x}+k_{y})+\cos(k_{x}-k_{y}) \right\} \nonumber \\ 
&+&t_{ss}^{3}\left( \cos 2k_{x}+\cos 2k_{y} \right) \nonumber \\ 
&+&t_{ss}^{4}\left\{ \cos(2k_{x}+k_{y})+\cos(2k_{x}-k_{y}) \right. \nonumber \\ 
& &+ \left. \cos(2k_{y}+k_{x})+\cos(2k_{y}-k_{x}) \right\} \nonumber \\ 
&+&t_{ss}^{5}\left\{  \cos(2k_{x}+2k_{y})+\cos(2k_{x}-2k_{y}) \right\} \nonumber \\ 
&+& E_0 ,
\end{eqnarray}
where $t_{ss}^{i}$ is the $i$-th nearest $s$-$s$ hopping integral.
We set $(t_{ss}^{1},t_{ss}^{2},t_{ss}^{3},t_{ss}^{4},t_{ss}^{5})
=(-0.5,-0.889,0.292,-0.229,0.687)$, and $E_0=1.33$.

Next, we explain the hybridization term.
Based on the Slater-Koster tight-binding method, 
the $s$-$f$ hybridization between the nearest Ce-sites is
\begin{eqnarray}
V_{\k f_{1}\uparrow}&=&-A_{1}t_{sf}(\sin k_{y} - i\sin k_{x} ),
\nonumber \\
V_{\k f_{2}\uparrow}&=&-A_2 t_{sf}(\sin k_{y} + i\sin k_{x} ),
\label{eqn:S-hybri}
\end{eqnarray}
and $V_{\k f_{l}\downarrow}= -V_{\k f_{l}\uparrow}^{*}$.
Here, $t_{sf}=(sf\sigma)$, and
$A_1=\sqrt{{18}/{14}}$ and $A_2=\sqrt{{3}/{7}}$.
Since $A_1>A_2$, the relation $D_{f_1}(0) > D_{f_2}(0)$ holds 
in the present two-dimensional PAM,
where $D_{f_l}(0)$ is the $f_l$-electron density-of-states
at Fermi level.
However, $D_{f_1}(0)=D_{f_2}(0)$ holds in the cubic model,
since the $s$-$f$ hybridization along $z$-axis 
is larger for $f_2$-electron.
To escape from the artifact of two-dimensionality,
we put $A_1=A_2=\sqrt{{18}/{14}}$ in the present study.

In the present $\Gamma_8$ model,
the relation $V_{\k f_{1}\s} \propto V_{\k f_{2}\s}^{*}$ holds
as shown in Eq. (\ref{eqn:S-hybri}).
In contrast, in the $\Gamma_7^{(1)}$-$\Gamma_7^{(2)}$ model
for CeCu$_2$Si$_2$ used in Ref. \cite{S-Tazai-HF},
the relation $V_{\k f_{1}\s} \propto V_{\k f_{2}\s}$ holds.

Finally, we explain the Coulomb interaction in $f$-electrons,
which is derived from Slater-Condon parameter $F^{p}$ 
\cite{S-Tazai-HF}.
We set $(F^{0},F^{2},F^{4},F^{6})=(5.3,9.09,6.927,4.756)$ in unit eV 
by referring Ref.\cite{S-F0F2F4F6}. 
The derived Coulomb interaction is about 6eV.
If we use the such large Coulomb interaction in the RPA,
the magnetic order appears 
since the self-energy is dropped in the RPA.
Therefore, we introduce the following Coulomb interaction term:
\begin{eqnarray}
&&\hat{H}_{U}=u \hat{H}^{0}_{U}, 
\\
&&\hat{H}^{0}_{U}=\frac14 \sum_{LL'MM'} U^0_{L,L';M,M'}
f_{L}^\dagger f_{L'} f_{M} f_{M'}^\dagger,
\end{eqnarray}
where $L=(l,\sigma)$ and $M=(m,\rho)$.
$u$ is the interaction model parameter, and 
$\hat{U}^0$ is the normalized Coulomb interaction
introduced in Ref. \cite{S-Tazai-HF}.
That is, the maximum element of $\hat{U}^0$ is
normalized to unity.

\subsection{B: multipole-operator}

Here, we list the pseudo-spin representation 
of the multipole operators in TABLE {\ref{tab:multipole},
which was first introduced in Ref.\cite{S-Shiina1}.
An even-rank (odd-rank) operator corresponds to 
an electric (magnetic) multipole operator.
Each multipole operator of rank $k$ are composed of  
$4\times 4$ tensor $J^{(k)}_{q} (q=-k\sim k)$
\cite{S-Shiina1,S-Springer} which is given by 
$ [J_{\pm},J^{(k)}_{q}]=\sqrt{(k\mp q)(k\pm q+1)}J^{(k)}_{q\pm1}$
$J_{k}^{(k)}=(-1)^{k}\sqrt{(2k-1)!!/(2k)!!}J_{+}^{k}$.
The multipole operators $\hat{Q}$
is given by the linear combination of $J^{(k)}_{q}$.
The $4\times4$ matrix form of each
electric (odd-rank) multipole operators is given by
\cite{S-Shiina1}
\begin{eqnarray}\Gamma_{1}^{+}&&
\begin{cases}
\hat{1}&=\hat{\sigma}^{0}\hat{\tau}^{0}  \nonumber \\
\hat{O}_{20}&=4.0\hat{\sigma}^{0} \hat{\tau}^{z} \nonumber \\
\end{cases} \nonumber \\ \Gamma_{3}^{+}&&
\begin{cases}
\hat{O}_{22}&= 4.0 \hat{\sigma}^{0} \hat{\tau}^{x}
\end{cases} \nonumber \\ \Gamma_{4}^{+}&&
\begin{cases}
\hat{O}_{xy}&=-\hat{\sigma}^{z} \hat{\tau}^{y}  \nonumber \\
\end{cases} \nonumber \\ \Gamma_{5}^{+}&&
\begin{cases}
\hat{O}_{yz}&=-\hat{\sigma}^{x} \hat{\tau}^{y}  \nonumber \\
\hat{O}_{zx}&=-\hat{\sigma}^{y} \hat{\tau}^{y}  \nonumber \\
\end{cases} \label{eqn:S-eleO}\\
\end{eqnarray}
The $4\times4$ matrix form of each
magnetic (odd-rank) multipole operators is given by
\cite{S-Shiina1}
\begin{eqnarray} \Gamma_{2}^{-}&&
\begin{cases}
\hat{J}^{z}&= \hat{\sigma}^{z} \left(-1.2\hat{\tau}^{0}-0.67 \hat{\tau}^{z} \right)  \nonumber \\
\hat{T}^{z\a}&= \hat{\sigma}^{z} \left(-1.0\hat{\tau}^{0}-7.0 \hat{\tau}^{z} \right)  \nonumber \\
\end{cases} \nonumber \\ \Gamma_{3}^{-}&&
\begin{cases}
\hat{T}^{xyz}&= -10.0 \hat{\sigma}^{0} \hat{\tau}^{y}  \nonumber \\
\end{cases}   \nonumber \\ \Gamma_{4}^{-}&&
\begin{cases}
\hat{T}^{z\b}&= -6.7 \hat{\sigma}^{z} \hat{\tau}^{x}  \nonumber \\
\end{cases}  \nonumber \\ \Gamma_{5}^{-}&&
\begin{cases}
\hat{J}^{x}&=\hat{\sigma}^{x} \left(1.2\hat{\tau}^{0}-0.34\hat{\tau}^{z} +0.58\hat{\tau}^{x}\right)  \nonumber \\
\hat{J}^{y}&=\hat{\sigma}^{y} \left(1.2\hat{\tau}^{0}-0.34\hat{\tau}^{z} -0.58\hat{\tau}^{x}\right)    \nonumber \\
\hat{T}^{x\a}&=  \hat{\sigma}^{x}  \left(\hat{\tau}^{0}-3.5\hat{\tau}^{z} +6.1\hat{\tau}^{x}\right)   \nonumber \\
\hat{T}^{y\a}&=  \hat{\sigma}^{y}  \left(\hat{\tau}^{0}+3.5\hat{\tau}^{z} +6.1\hat{\tau}^{x}\right)   \nonumber \\
\hat{T}^{x\b}&=  \hat{\sigma}^{x}  \left(-5.8\hat{\tau}^{z}-3.4\hat{\tau}^{x}\right)   \nonumber \\
\hat{T}^{y\b}&=  \hat{\sigma}^{y}  \left(-5.8\hat{\tau}^{z}+3.4\hat{\tau}^{x}\right)  
\end{cases} \label{eqn:S-magneO} \\
\end{eqnarray}
%

In the main text, we use the normalized multipole matrix
introduced as follows:
\begin{eqnarray}
\hat{Q} /\sqrt{\sum_{L,M}|Q_{L,M}|^2} 
\rightarrow \hat{Q}  .
\end{eqnarray}
Then, the normalized $\hat{Q}$ 
satisfies the condition $\sum_{L,M}|Q_{L,M}|^2=1$.

\subsection{C: multipole fluctuations}

In the main text, we explain the analytic expression 
only for AL1 term.
The expression for the AL2 term is given as
\begin{eqnarray}
X^{\rm{AL2}}_{\a \b}(q)&=\frac{T}{2}\sum_{\a' \b' \a'' \b''}
\Lambda_{\a' \b''}^{\a} (q,p) V_{\b'' \b'} (p-q) \nonumber \\
&\times  V_{\a'' \a'}(p) \tilde{\Lambda}_{\a'' \b'}^{\b} (q,p),
\label{eqn:S-UALc}
\end{eqnarray}
where 
\begin{eqnarray}
\Lambda^{EF}_{ABCD} (q,p) \equiv -T\sum_{k}G^{f}_{BF}(k-q)G^{f}_{ED}(k)G^{f}_{CA}(k-q+p),
\nonumber \\
\tilde{\Lambda}^{EF}_{ABCD} (q,p) \equiv -T\sum_{k}G^{f}_{AE}(k+q)G^{f}_{FC}(k)G^{f}_{DB}(k+q-p).
\nonumber
\end{eqnarray}
The expression for the MT term is 
\begin{eqnarray}
&& X^{MT}_{LL'MM'}(q)=T^{2}\sum_{p,k,A\sim D}
G_{LA}(k+q-p)G_{BL'}(k-p)\nonumber \\ 
&&\hspace{5pt} \times G_{DM}(k+q)G_{M'C}(k)V_{DACB} (p) .
\label{eqn:S-UALMT}
\end{eqnarray}
The total VC is given by
$\hat{X}^{\rm AL+MT}=\hat{X}^{AL1}+\hat{X}^{AL2}+\hat{X}^{MT}$,
by subtracting the double counting 
second order diagrams of order $u^2$.

\begin{figure}[htb]
\vspace{5mm}
\includegraphics[width=.8\linewidth]{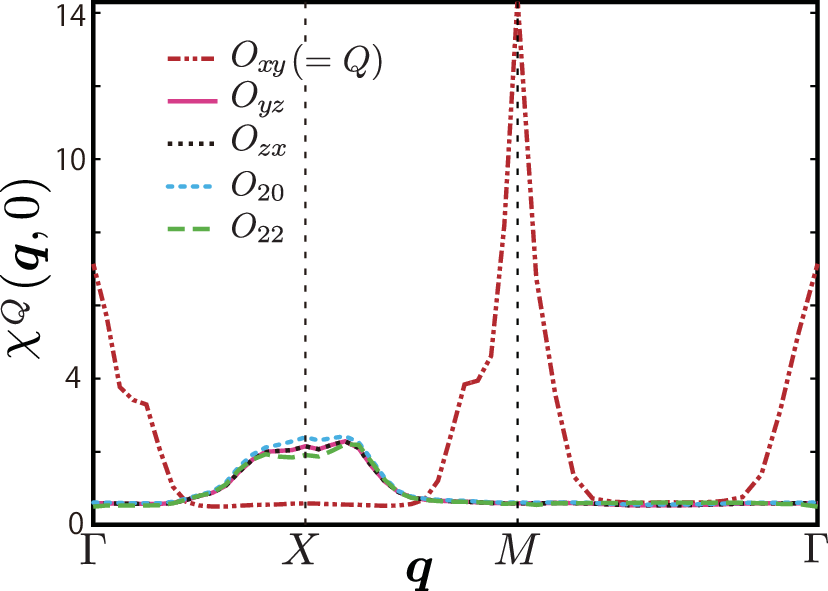}
\caption{
Obtained quadrupole susceptibilities $\chi^Q(\q,0)$
for $Q=O_{xy}$, $O_{zx/yz}$, and $O_{20/22}$.
}
\label{fig:S-suscep}
\end{figure}

In the main text, 
we perform the numerical study of multipole susceptibilities
by considering both MT- and AL-VCs, and 
showed that $O_{xy}$ octupole susceptibility 
is strongly enlarged by the AL-VCs.
Here, we show all the quadrupole susceptibilities
obtained by the present study in Fig. \ref{fig:S-suscep}.
In the cubic model,
$\chi^{Q}(\q,0)$ with $Q=O_{xy},O_{zx},O_{yz}$
should equally develop.
In the present two-dimensional model, however,
only $O_{xy}$-fluctuation strongly develops.
The reason is that ($T_{x},T_{y}$) fluctuations
are much larger than $T_{z}$ fluctuations in the RPA,
due to the violation of cubic symmetry.
Since $O_{\mu\nu}$ quadrupole susceptibility is magnified by 
($T_{\mu},T_{\nu})$ fluctuations ($\mu,\nu=x,y,z$)
due to the AL-VC,
$\chi^{O_{xy}}(\q,0)$ is the largest in the present model.

As we show in TABLE \ref{tab:multipole2},
the Coulomb interaction $U^{Q}_0$ for $Q=O_{xy/yz/zx}$
is much larger than that for $Q=O_{20/22}$.
For this reason, it is difficult to expect that 
$Q=O_{20/22}$ quadrupole susceptibility becomes 
larger than $Q=O_{xy}$ one, even if the AL-VCs are considered.
Thus, the relation $\chi^{O_{xy}}(\q,0) > \chi^{O_{20/22}}(\q,0)$
should hold even in cubic systems.

\begin{figure}[htb]
\vspace{5mm}
\includegraphics[width=.8\linewidth]{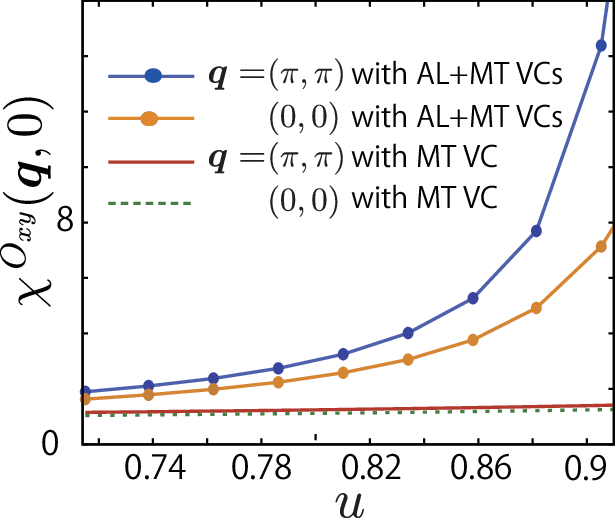}
\caption{
Obtained 
$\chi^{O_{xy}}_{\rm AL}(\q,0)$ with AL1+AL2 terms
and $\chi^{O_{xy}}_{\rm MT}(\q,0)$ with MT term
at $\q=\bm{Q},\bm{0}$ as function of $u$.
}
\label{fig:S-ALMT}
\end{figure}

Next, we calculate the susceptibility with AL-VC (MT-VC),
$\chi^{O_{xy}}_{\rm{AL(MT)}}(\q,0)$, given by 
$\hat{\phi}(q)=\hat{\chi}^{0}(q)+\hat{X}^{\rm{AL(MT)}}(q)$.
Figure \ref{fig:S-ALMT} shows the obtained
$\chi^{O_{xy}}_{\rm{AL}}(\q,0)$ and $\chi^{O_{xy}}_{\rm{MT}}(\q,0)$
as functions of $u$.
$\chi^{O_{xy}}_{\rm{AL}}(\q,0)$ strongly increases with $u$,
similarly to $\chi^{O_{xy}}(\q,0)$ with AL+MT terms
shown in Fig. 3 (c) in the main text.
In contrast, $\chi^{O_{xy}}_{\rm{MT}}(\q,0)$ 
remains small and comparable to the RPA result in Fig. 3 (c).
Therefore, it is verified that the enhancement of 
$O_{xy}$ quadrupole fluctuations originates from the AL-VC,
whereas the MT-VC is very small.

To understand this result analytically,
we analyze the AL and MT terms for the electric multipole channel
given by the following magnetic multipole susceptibility 
\begin{eqnarray}
\chi^{\rm mag}(\q,\w_l)&=& \frac{a\xi^2}{1+\xi^2(\q-\Q)^2+|\w_l|/\w_{\rm mag}},
\end{eqnarray}
where $\xi^2\propto (T-T_0)^{-1}$ and $\w_{\rm mag} \propto \xi^{-2}$.
$\xi$ is the correlation length.
Then, in two-dimensional systems at a fixed $T$,
AL-VC and MT-VC given in Eqs. (\ref{eqn:S-UALc})-(\ref{eqn:S-UALMT}) 
are scaled as
$X^{\rm AL}(\bm{0},0)\sim \sum_\p\{\chi^{\rm mag}(\p,0)\}^2\sim \xi^2$
and 
$X^{\rm MT}(\bm{0},0)\sim \sum_\p\chi^{\rm mag}(\p,0)\sim \log\xi$,
respectively.
Therefore, the AL term dominates over the MT term when $\xi\gg1$
\cite{S-Onari}.
The significance of the AL terms
near the magnetic criticality
is verified by the functional-renormalization-group (fRG) study
\cite{S-RG1,S-RG2,S-RG3}.

In $d$-dimensional system,
the AL term is proportional to $\max\{\xi^{4-d},1\}$.
This fact means that the non-locality of irreducible AL diagram
is significant near the magneitc criticality $(\xi\gg1)$.



\begin{thebibliography}{99}
\bibitem{comment2}
At $T\sim T_Q$, 
$\chi^{J_\mu}(\q,\w)$ at $\q=\bm{Q},\bm{0}$
is much larger than that 
at the magnetic order wavevector $\bm{Q}_N=(\pi/2,\pi/2,0)$
below $T_N$ according to Ref. \cite{Inosov2}.
\bibitem{comment}
We adjust $u$ to keep $\a^{\rm mag}$ constant, since
$\a^{\rm mag}$ decreases with $h_z$ for fixed $u$ in the RPA.
In the FLEX approximation, $\a^{\rm mag}$ increases with $h_z$
due to the negative feedback effect between
spin fluctuations and self-energy
\cite{Sakurazawa}.
\end{thebibliography}
\end{document}